\documentclass{article}
\PassOptionsToPackage{numbers,compress}{natbib}
\usepackage[final,main]{neurips_2026}

\usepackage[utf8]{inputenc}
\usepackage[T1]{fontenc}
\usepackage{microtype}
\usepackage{graphicx}
\usepackage{subcaption}
\usepackage{booktabs}
\usepackage{multirow}
\usepackage{makecell}
\usepackage{array}
\usepackage{tabularx}
\usepackage{adjustbox}
\usepackage[table]{xcolor}
\usepackage{amsmath}
\usepackage{amssymb}
\usepackage{amsfonts}
\usepackage{nicefrac}       
\usepackage{amsthm}
\usepackage{bbm}
\usepackage{pifont}
\usepackage{mathtools}
\usepackage{algorithm}
\usepackage{algorithmic}
\usepackage{enumitem}
\usepackage{wrapfig}
\usepackage{hyperref}       %
\usepackage[capitalize,noabbrev]{cleveref}
\usepackage{url}
\usepackage{xspace}
\usepackage{caption}

\newcommand{\wh}{\emph{watermarking hallucination}\xspace}

\theoremstyle{plain}

\theoremstyle{definition}

\title{Invisible Ink, Visible Lies: How Production Watermarking Causes LLMs to Hallucinate}

\author{
\textbf{Haocheng Ye}$^{1}$\thanks{Equal contribution.}
\quad
\textbf{Aoting Hu}$^{2}$\footnotemark[1]
\quad
\textbf{Xinwei Zhang}$^{3}$\thanks{Corresponding authors.}
\\[0.5ex]
\textbf{Xunzhu Tang}$^{4}$
\quad
\textbf{Shuchao Pang}$^{1}$\footnotemark[2]
\quad
\textbf{Jason Xue}$^{5}$
\\[1ex]
$^{1}$Nanjing University of Science and Technology
\\
$^{2}$Anhui University of Technology
\\
$^{3}$The Hong Kong Polytechnic University
\\
$^{4}$University of Luxembourg
\\
$^{5}$CSIRO
}

\begin{document}

\maketitle

\begin{abstract}
Text watermarking helps identify AI-generated content, but its effect on factual reliability remains underexplored.
In this paper, we study \emph{watermarking hallucination}: factual errors induced or amplified by watermarking even when the required evidence is present in the context and the unwatermarked model can answer correctly.
Using a controlled retrieval-augmented generation setting, we compare unwatermarked and watermarked generations under the same context, query, and decoding configuration, and quantify their factual accuracy decrease.
Across six representative watermarking methods, including KGW, SWEET, DiPmark, GumbelSoft, Gumbel-Max, and SynthID watermarking, we consistently observe watermark-induced hallucination. Watermarked outputs can remain fluent while introducing factual errors.
We attribute this failure mode to two mechanisms: \ding{172} token perturbations in the current-step arising from method-specific reweighting or keyed sampling, and \ding{173} prefix-induced attention drift, which accumulates through autoregressive decoding and weakens later attention to the factual context.
Motivated by this analysis, we propose two plug-in interventions at the token and attention levels that can be integrated into existing watermarking methods to improve factuality.
At a matched TPR of 0.90 at 1\% FPR, combining the two interventions reduces factual errors by approximately 90\% relative to watermark-only decoding while preserving fluency and comparable decoding efficiency.
Overall, this work highlights factuality as a first-class criterion in watermark evaluation, alongside detectability and robustness, and calls for careful factuality validation before deploying watermarks in fact-critical applications.
Code is available at \url{https://github.com/Haocheng-NJUST/WM-Hallucination}.
\end{abstract}

\section{Introduction}

Large language models (LLMs) increasingly generate long-form text in settings where provenance and factual reliability both matter, including clinical notes, legal summaries, financial reports, and public-facing assistants~\cite{10.1145/3442188.3445922,mirsky2023offensive,dahl2024legalfictions,https://doi.org/10.1111/jels.12413}. Text watermarking offers a way to identify model-generated content by embedding a statistical signal into the decoding process~\cite{pmlr-v202-kirchenbauer23a,kuditipudi2024robust,Dathathri2024ScalableWF}. Some watermarking methods explicitly reweight token probabilities, while others preserve the model distribution under prescribed randomization. In both cases, the realized watermark-conditioned token choices can alter generation trajectories and factual outcomes.

\begin{figure}[t]
\centering
\begin{subfigure}{0.48\linewidth}
  \centering
  \includegraphics[width=\linewidth]{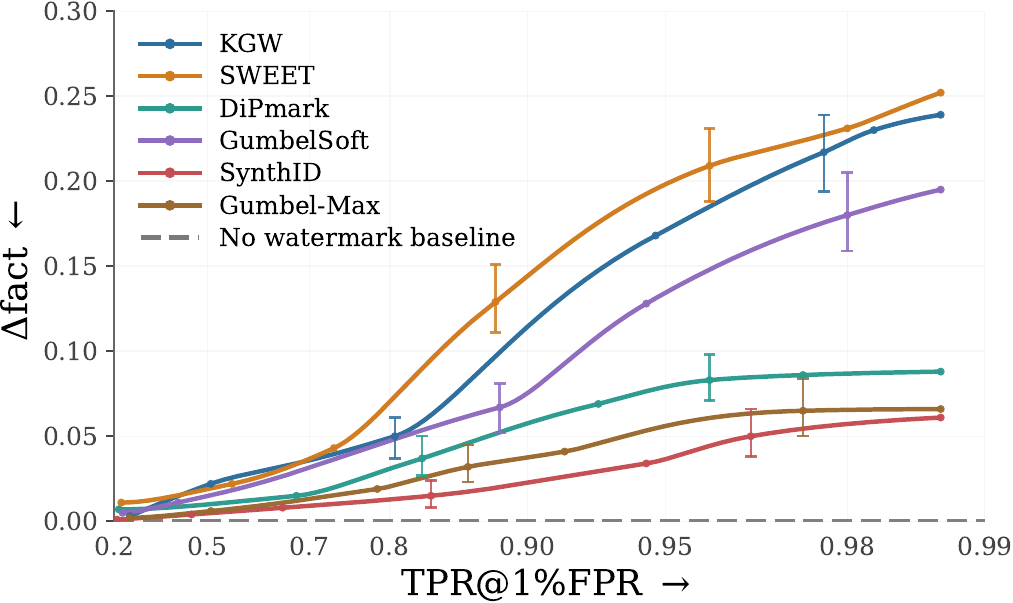}
  \caption{Factual accuracy decrease across methods.}
\end{subfigure}
\hfill
\begin{subfigure}{0.48\linewidth}
  \centering
  \includegraphics[width=\linewidth]{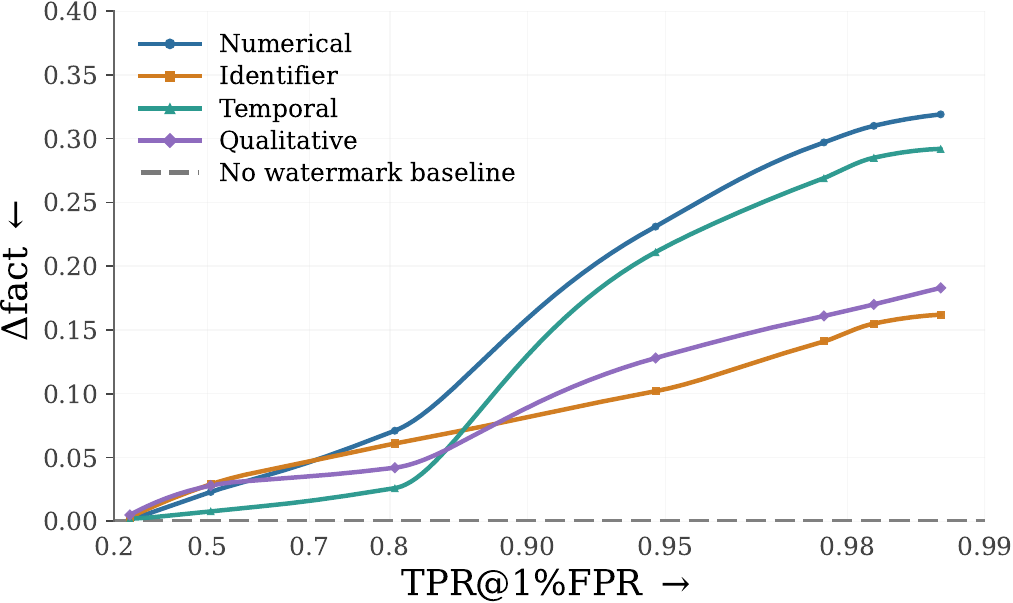}
  \caption{Fact-type breakdown for KGW~\cite{pmlr-v202-kirchenbauer23a}.}
\end{subfigure}
\caption{
\textbf{Factual accuracy decrease under watermarking.}
(a) Net factual accuracy decrease, \(\Delta_{\mathrm{fact}}\),
as a function of watermark detection true-positive rate (TPR)
at a fixed false-positive rate (FPR) of 1\%, across six watermarking methods.
Error bars indicate paired bootstrap confidence intervals.
(b) Breakdown of \(\Delta_{\mathrm{fact}}\) by fact type for KGW.
}
\label{fig:WHR_ALL}
\end{figure}

Existing watermark evaluations typically assess watermarked text from two perspectives: \textit{watermark effectiveness}, measured by detectability and robustness, and \textit{output quality}, measured by perplexity, semantic similarity, and readability~\cite{liu2024surveytextwatermarking,yang2025watermarksurvey,NEURIPS2024_21b5883b}. These criteria are necessary for deployment, but they often ignore whether the watermarked output preserves facts supported by the prompt or retrieved context. Recently, \Citet{hastuti-etal-2025-factuality} showed that watermarked medical text can remain fluent while changing a dosage, date, identifier, or legal citation. They show that factuality can degrade even when coherence-oriented quality measures remain acceptable. However, their study does not isolate the factual degradation caused by watermarking itself, explain why such errors arise during decoding, or propose a mitigation strategy.

We refer to this failure mode as \wh and study it in a controlled retrieval-augmented generation (RAG) setting~\cite{gao2024retrievalaugmentedgenerationlargelanguage,ni2025trustworthyretrievalaugmentedgeneration}. Although retrieval and grounding can reduce LLM hallucinations~\cite{10.1145/3703155,xu-etal-2024-knowledge-conflicts,tao-etal-2024-context}, watermarking can introduce factual errors even when the relevant facts are available in the supplied context and the same model answers correctly without watermarking. We quantify this effect by the net factual accuracy decrease, \(\Delta_{\mathrm{fact}} = \mathrm{Acc}_{\mathrm{clean}} - \mathrm{Acc}_{\mathrm{wm}}\), where \(\mathrm{Acc}_{\mathrm{clean}}\) and \(\mathrm{Acc}_{\mathrm{wm}}\) denote factual accuracy for paired unwatermarked and watermarked outputs, respectively, evaluated under matched conditions. Positive values indicate lower factual accuracy under watermarking. Our evaluation covers six watermarking methods: KGW~\cite{pmlr-v202-kirchenbauer23a}, SWEET~\cite{lee-etal-2024-wrote}, DiPmark~\cite{wu2024resilientaccessibledistributionpreservingwatermark}, GumbelSoft~\cite{fu-etal-2024-gumbelsoft}, SynthID-style watermarking~\cite{Dathathri2024ScalableWF}, and Gumbel-Max~\cite{kuditipudi2024robust}. As shown in \Cref{fig:WHR_ALL}(a), factual accuracy decreases across the evaluated methods, with larger decreases generally observed at higher watermark detection rates. The KGW breakdown in \Cref{fig:WHR_ALL}(b) further shows that numerical and temporal facts exhibit larger decreases than the other fact types considered.

To investigate the cause of this degradation, we compare factual accuracy between watermarked and non-watermarked generation, and analyze output logits and internal attention patterns. We identify two coupled pathways. At the token level, distortionary methods can directly bias the decoder against a fact-consistent token. At the prefix level, earlier watermark-conditioned token choices can alter the generated prefix and generation trajectory, thereby reducing subsequent attention to fact-bearing context. This \emph{prefix-induced drift} is more difficult to address and has broader consequences than a local token substitution, because even non-factual prefix changes can accumulate and increase the likelihood of later factual errors.

Motivated by these findings, we introduce two fact-preserving interventions, which can be integrated as plug-in components into existing watermarking methods to improve the factual accuracy of watermarked text. Specifically, Fact-Preserving Token Intervention (FPTI) limits watermark-conditioned changes at factual positions, whereas Fact-Preserving Attention Intervention (FPAI) targets prefix-mediated attention loss by restoring attention toward fact-bearing context. At TPR@1\%FPR \(=0.90\), their joint application reduces \(\Delta_{\mathrm{fact}}\) by around 90\% relative to watermark-only decoding, substantially improving the trustworthiness of watermarked generation. Nevertheless, a residual gap remains under strong watermarking. This gap reflects a hard constraint: even when a method preserves the model distribution under its prescribed randomization, a realized watermark-conditioned trajectory can still diverge from the narrow path required by exact factual grounding.

Overall, the paper makes three contributions.
\begin{itemize}[leftmargin=*]
    \item We identify \wh as a factual reliability failure mode distinct from surface-quality degradation and ordinary hallucination, and quantify watermark-induced factual loss under paired controlled RAG evaluation.
    
    \item We attribute watermarking hallucination to two coupled decoding-level pathways: token-level perturbations, arising through systematic bias in distortionary methods, and prefix-induced attention interference, which weakens later reliance on factual context.
    
    \item We propose two complementary interventions that protect factual token selection and restore attention to supporting evidence, reducing factual degradation under watermarking across multiple watermarking methods and model families.
\end{itemize}

\section{Problem Formulation}
\label{sec:problem_formulation}

\subsection{Context-Grounded Generation}

We study factual generation under a context-grounded setting. Let \(x\) denote
the user prompt together with the retrieved or provided factual context. The
model generates an answer \(y_{1:T}\) autoregressively:
\begin{equation}
\label{eq:base-generation}
    p_\theta(y_{1:T}\mid x)
    =
    \prod_{t=1}^{T}
    p_\theta(y_t\mid x,y_{<t}).
\end{equation}
At each decoding step \(t\), the model assigns a token sampling score
\(z_t(y_t\mid x,y_{<t})\) to each candidate token \(y_t\). In a factual task, some
answer spans are expected to be supported by the context. We call these spans
\emph{factual decision points}. At such a point, the model must select a token
that is consistent with the evidence in \(x\), such as an entity name, date,
number, identifier, or relation.

In standard generation, factual errors may arise when the next-token distribution is not sufficiently anchored to the relevant evidence. This can
happen because of missing parametric knowledge, conflict between model memory
and the provided context, or gradual drift away from the evidence during
generation~\cite{10.1145/3703155,xu-etal-2024-knowledge-conflicts,tao-etal-2024-context}. In
RAG-based generation, the supplied context is expected to reduce this
uncertainty. Therefore, factuality depends not only on whether the final string
looks fluent, but also on whether the relevant evidence remains active when
each factual span is generated.

\subsection{Decoding-Time Watermarking}

We focus on decoding-time watermarking, which modifies token selection
during generation without changing the model parameters~\citep{pmlr-v202-kirchenbauer23a,lee-etal-2024-wrote,wu2024resilientaccessibledistributionpreservingwatermark,fu-etal-2024-gumbelsoft}. Let
\(\tilde{y}_{1:T}\) denote the sequence generated under watermarking. At step
\(t\), the model computes the token sampling scores
\(z_t(y_t\mid x,\tilde{y}_{<t})\) over candidate tokens \(y_t\),
conditioned on the watermarked prefix \(\tilde{y}_{<t}\).
The watermarking operator \(\mathcal{W}_t\) transforms these into watermarked scores:
\begin{equation}
\label{eq:wm-distribution}
    \tilde{z}_t
    =
    \mathcal{W}_t\!\left(
    z_t(\cdot\mid x,\tilde{y}_{<t}),
    \tilde{y}_{<t}
    \right).
\end{equation}
Here, \(\cdot\) denotes scores over the full vocabulary; dependence on the
watermark key and sampling randomness is omitted for brevity.
We define the resulting token-level score perturbation as
\begin{equation}
\label{eq:wm-logit}
    b_t(y_t;\tilde{y}_{<t})
    =
    \tilde{z}_t(y_t)
    -
    z_t(y_t\mid x,\tilde{y}_{<t}).
\end{equation}

The next token is selected from the watermarked scores using the
method-specific sampling rule. These scores depend on a prefix
already shaped by earlier watermarking steps, rather than on the
prefix of an unwatermarked generation. Watermarking therefore
affects the next token both by changing its current sampling score
and by changing the preceding tokens on which that score depends.

\subsection{Watermarking Hallucination}
\label{sec:watermarking_hallucination}

We define \emph{watermarking hallucination} as factual degradation induced or amplified by the watermarking process.
Let \(F(\cdot,x)\) be a factuality evaluator that checks whether an answer is supported by the context \(x\), where \(F(y,x)=1\) denotes a factual answer and \(F(y,x)=0\) denotes a factual error.
For a paired generation under the same prompt and decoding configuration, let \(y^0\) be the unwatermarked output and \(\tilde{y}\) be the watermarked output.
A watermarking hallucination occurs when the base generation is factual but the corresponding watermarked generation is not:
\begin{equation}
\label{eq:wh-definition}
    F(y^0,x)=1,
    \qquad
    F(\tilde{y},x)=0 .
\end{equation}

This paired definition isolates factual errors introduced when watermarking shifts an otherwise context-supported generation toward a factually inconsistent alternative.
Since watermarking may also turn an incorrect base generation into a factual one, we quantify its aggregate effect using the \emph{factual accuracy decrease}:
\begin{equation}
\label{eq:delta-fact}
\begin{aligned}
\Delta_{\mathrm{fact}}
={}&
\frac{1}{|\mathcal{D}|}
\sum_{(x,q)\in\mathcal{D}}
\Big(
\mathbb{I}\!\left[
F(y^0,x)=1 \wedge F(\tilde{y},x)=0
\right]
\\[-2pt]
&\qquad\qquad -
\mathbb{I}\!\left[
F(y^0,x)=0 \wedge F(\tilde{y},x)=1
\right]
\Big).
\end{aligned}
\end{equation}
The first term counts answers that change from correct to incorrect
under watermarking, while the second counts those that change
from incorrect to correct.
Their difference, \(\Delta_{\mathrm{fact}}\), measures the net
factual accuracy decrease relative to unwatermarked generation.
Positive values indicate degradation, and negative values indicate improvement.

\begin{figure*}[t]
\centering
\includegraphics[width=\linewidth]{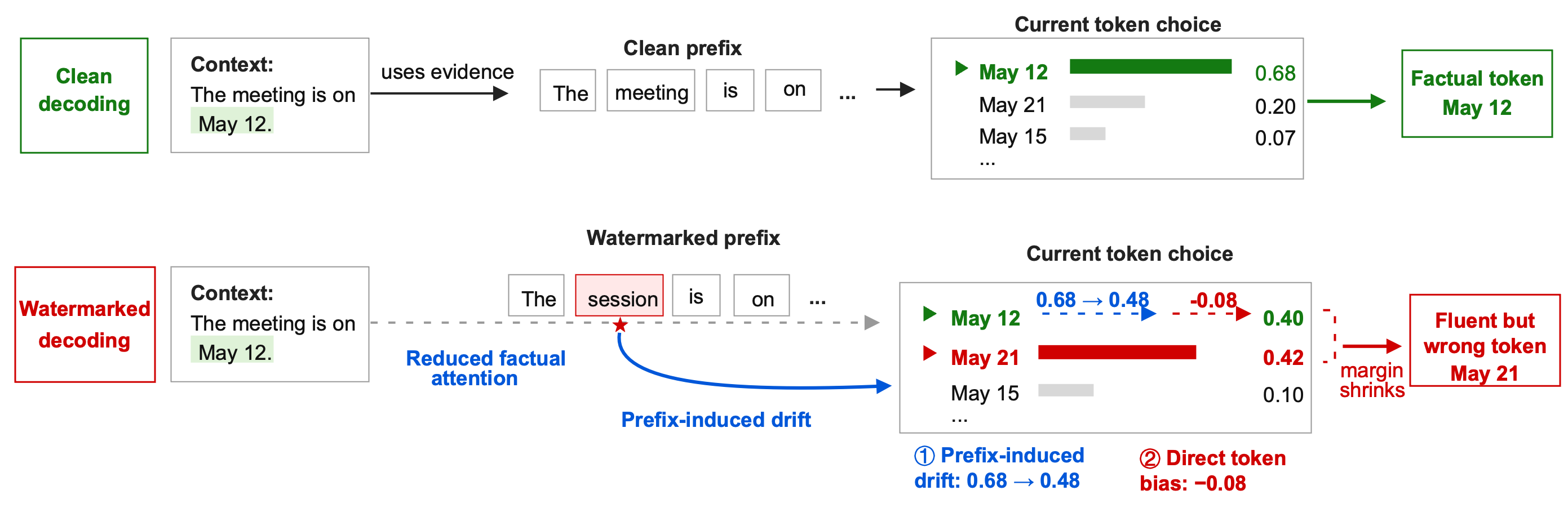}
\caption{
\textbf{Decomposition of watermarking hallucination.}
Clean decoding preserves the factual margin by following the fact-bearing context.
Watermarked decoding can first induce prefix drift, e.g., changing ``meeting'' to the fluent alternative ``session'', which weakens factual reliance and lowers the score of the factual token.
The current-step score perturbation then further compresses the margin, allowing a fluent distractor such as \emph{May 21} to overtake the evidence-supported token \emph{May 12}.
}
\label{fig:mechanism_overview}
\end{figure*}

\section{Mechanism Analysis}
\label{sec:mechanism_analysis}

\subsection{A Decomposition of Watermarking Hallucination}
\label{sec:mechanism_margin_drift}

Watermarking hallucination can be understood as a recursive compression of the factual margin.
At a factual decision point, let \(y_t^\star\) denote the fact-consistent token supported by the context, and let \(V_t\) be the effective candidate set under the sampling rule.
Under the clean generation trajectory \(y^0_{<t}\), we define the factual robustness margin as
\begin{equation}
\label{eq:margin}
    \epsilon_t(y_t^\star)
    =
    z_t(y_t^\star\mid x,y^0_{<t})
    -
    \max_{y_t\in V_t\setminus\{y_t^\star\}}
    z_t(y_t\mid x,y^0_{<t}).
\end{equation}
This margin measures how far the evidence-supported token leads its strongest plausible alternative before watermarking.

As illustrated in \Cref{fig:mechanism_overview}, watermarking can erode this margin through two coupled effects.
First, earlier watermark-conditioned token choices become part of the prefix and shift the model's later distribution.
Even if such choices are fluent and semantically close to the clean prefix, they can weaken reliance on the original fact-bearing context.
Second, the current watermarking rule can alter the effective selection scores of candidate tokens or the realized token selection, favoring a fluent but factually incorrect competitor over the evidence-supported token.
Distortionary methods systematically reweight token probabilities.
Distribution-preserving methods preserve the marginal distribution
but can still change individual token choices.
Thus, factual errors can arise when prefix-induced drift and the current-step token perturbation jointly reverse the local margin.

Formally, let \(y^0_{1:T}\) and \(\tilde{y}_{1:T}\) denote the clean and watermarked generation trajectories, and let \(\tilde{z}_t(y_t)\) denote the watermarked score defined above. The comparison below requires a common score representation and scale within each method.
At step \(t\), its change relative to clean decoding can be decomposed as
\begin{equation}
\label{eq:decomposition}
\begin{aligned}
    \tilde{z}_t(y_t)
    -
    z_t(y_t\mid x,y^0_{<t})
    &=
    \underbrace{
    b_t(y_t;\tilde{y}_{<t})
    }_{\text{\ding{172} token-level perturbation}}
    +
    \underbrace{
    z_t(y_t\mid x,\tilde{y}_{<t})
    -
    z_t(y_t\mid x,y^0_{<t})
    }_{\text{\ding{173} prefix-induced drift}} .
\end{aligned}
\end{equation}
The first term is the current-step score perturbation defined in \Cref{eq:wm-logit}; it does not necessarily imply a systematic bias in the marginal token distribution.
The second term captures the score change caused by conditioning on the watermarked prefix rather than the clean prefix.
This prefix-induced drift is recursive: a watermark-conditioned token changes the prefix, which shifts the sampling scores at subsequent steps.

This view explains why watermarking can introduce factual errors without causing visible incoherence.
The incorrect token may be locally fluent and semantically close to the correct one, but factual grounding often requires selecting a narrow set of exact entities, numbers, dates, and identifiers.
Watermarking hallucination arises when the current-step token perturbation, prefix-induced drift, or their combination erases the margin in \Cref{eq:margin}.

\subsection{Empirical Evidence of the Two Channels}
\label{sec:two_channels}

The decomposition in \Cref{eq:decomposition} suggests two channels through which watermarking affects factual generation.

\noindent\textbf{\ding{172} Direct token-level perturbation.}
At the current decoding step, watermarking can affect token selection through method-specific reweighting or keyed sampling rules.
Because the watermark signal is not explicitly fact-aware, a fact-consistent token may receive weaker support or lose the realized selection to a plausible but incorrect competitor.
This channel explains local substitutions in names, dates, numbers, and identifiers while preserving fluency.

\noindent\textbf{\ding{173} Prefix-induced attention interference.}
Prefix-induced drift captures the effect of conditioning on the watermarked prefix rather than the clean prefix.
In context-grounded generation, we measure one observable form of this drift by tracking how much the current decoding step attends to the fact-bearing tokens in the retrieved context.
Let \(\mathcal{F}(x)\subseteq\{1,\ldots,|x|\}\) denote the positions of input-side context tokens that support the target fact, and let \(A_t(j)\) be the attention weight from the current decoding position \(t\) to context token \(j\).
For example, in \Cref{fig:mechanism_overview}, \(\mathcal{F}(x)\) contains the context tokens corresponding to ``May 12'', because they provide the evidence for the next factual token.
We define the factual attention coefficient as
\begin{equation}
\label{eq:alpha}
    \alpha_t
    =
    \sum_{j\in \mathcal{F}(x)} A_t(j).
\end{equation}
Thus, \(\alpha_t\) measures attention to the supporting evidence in the RAG context.
A watermarked prefix such as ``The session is on'' can change the current decoding state and reduce attention to the context evidence ``May 12'', making the model more likely to follow parametric priors or fluent distractors.

Together, these two channels explain why watermarking hallucination can occur without visible incoherence: watermarking searches within the broad space of fluent continuations, whereas factual grounding often requires a narrow path through exact entities, numbers, dates, and relations.

\begin{figure}[t]
\centering
\begin{subfigure}{0.47\linewidth}
  \centering
  \includegraphics[width=\linewidth]{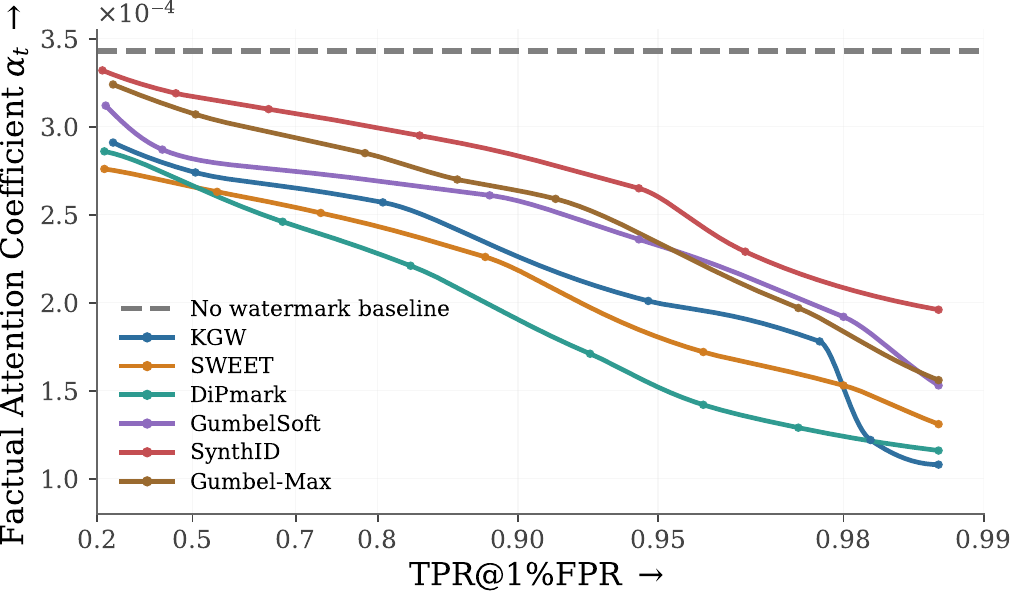}
  \caption{Attention to facts during watermarking}
\end{subfigure}
\hfill
\begin{subfigure}{0.45\linewidth}
\label{subfig:attention shift}
  \centering
  \includegraphics[width=\linewidth]{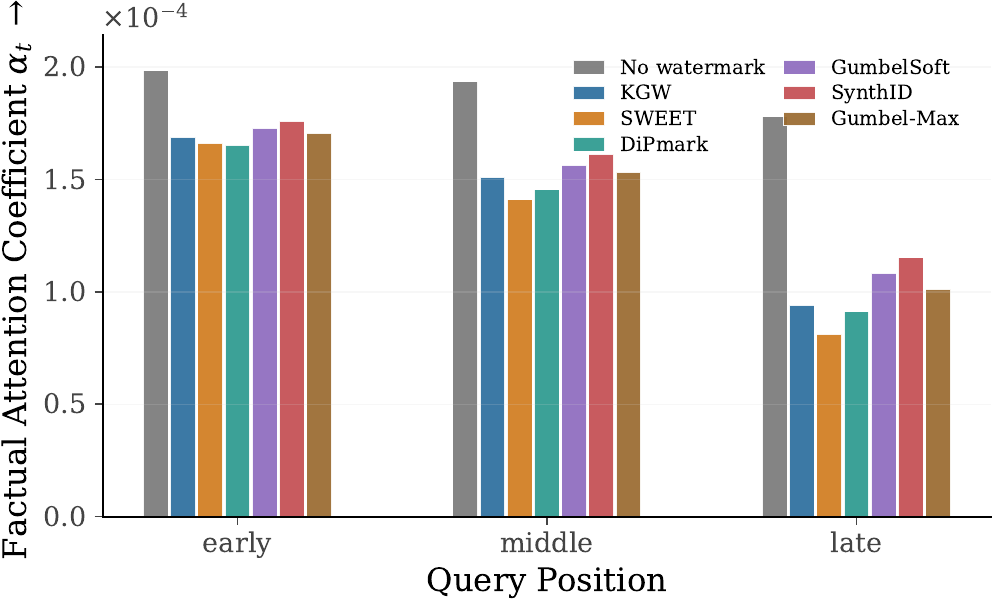}
  \caption{Watermarking hallucination avalanche}
\end{subfigure}
\caption{
\textbf{Attention-level interference under watermarking.}
(a) Factual attention coefficient $\alpha_t$ decreases as watermark detectability increases, measured by TPR@1\%FPR, across six representative methods.
(b) At a fixed detectability level, $\mathrm{TPR@1\%FPR}=0.9$, factual attention is lower when the same target query is answered later, indicating that attention degradation accumulates over the watermarked prefix.
}
\label{fig:attention_interference}
\end{figure}

\noindent\textbf{Empirical evidence.}
To isolate prefix-induced interference from the current-step token perturbation, we watermark ordinary answer tokens but disable the watermark perturbation when the next token is factual, i.e., \(b_t(y_t^\star;\tilde{y}_{<t})=0\).
We then measure \(\alpha_t\) over the supporting context tokens.
As shown in \Cref{fig:attention_interference}(a), \(\alpha_t\) consistently decreases as the detectability of the already generated prefix increases.
Since the factual token itself is not directly watermarked, this trend supports a prefix-mediated attention effect.

We further test whether this effect accumulates over longer generations by fixing the evidence block and target query while varying only the target query position:
\begin{equation}
\label{eq:avalanche-layout}
\begin{aligned}
P_{\mathrm{early}}  &= [\mathcal{C};\, Q^\star, Q_1, Q_2],\\
P_{\mathrm{middle}} &= [\mathcal{C};\, Q_1, Q^\star, Q_2],\\
P_{\mathrm{late}}   &= [\mathcal{C};\, Q_1, Q_2, Q^\star].
\end{aligned}
\end{equation}
All other factors, including the supporting evidence \(E^\star\), target query \(Q^\star\), prompt template, sampling parameters, and matched detectability level, are fixed.
\Cref{fig:attention_interference}(b) shows that factual attention is lower when the same query is answered later under watermarking, while the unwatermarked baseline shows a much smaller position effect.
This pattern indicates a \emph{watermark avalanche}: early watermark-conditioned choices alter the prefix, which weakens later attention to evidence and increases the risk of factual errors.
Implementation details for attention measurement and detectability alignment are provided in Appendix~\ref{sec:appendix_attention_detectability}.

\section{Fact-Preserving Interventions}
\label{sec:interventions}

Mechanism analysis identifies two sources of watermarking hallucination: current-step token perturbations induced by the watermarking rule and prefix-induced drift accumulated through autoregressive decoding.
We therefore introduce two targeted interventions.
Fact-Preserving Token Intervention (FPTI) neutralizes watermark perturbations on fact-bearing output tokens, while Fact-Preserving Attention Intervention (FPAI) compensates for the reduced attention to fact-bearing context.
Both interventions are designed as plug-in components for decoding-time watermarking methods: FPTI operates on output logits or sampling scores, whereas FPAI requires access to attention logits during inference.

\subsection{Fact-Preserving Token Intervention}
\label{sec:fpti}

FPTI targets the token-level perturbation term in \Cref{eq:decomposition}.
Given the input context \(x\), we construct a vocabulary-level factual token set
\(\mathcal{V}_{\mathrm{fact}}\subseteq\mathcal{V}\), which contains tokens
corresponding to entities, dates, numerical values, identifiers, and other
fact-bearing spans in the context.
In our implementation, these tokens are extracted automatically from the RAG
context~\citep{NEURIPS2020_6b493230} using named-entity recognition~\citep{ding-etal-2025-selecting}
and deterministic pattern-based matching~\citep{tokensregex2014}.

During decoding, FPTI removes the current-step watermark perturbation from factual tokens while leaving the watermarking rule unchanged elsewhere:
\begin{equation}
\label{eq:fpti}
\tilde{z}_t^{\mathrm{FPTI}}(y_t)
=
\begin{cases}
z_t(y_t\mid x,\tilde{y}_{<t}),
& y_t \in \mathcal{V}_{\mathrm{fact}},\\
z_t(y_t\mid x,\tilde{y}_{<t}) + b_t(y_t;\tilde{y}_{<t}),
& y_t \notin \mathcal{V}_{\mathrm{fact}}.
\end{cases}
\end{equation}
Here \(b_t(y_t;\tilde{y}_{<t})\) is the score perturbation defined in \Cref{eq:wm-logit}.
FPTI prevents this rule from directly changing the effective selection scores of tokens that carry factual content.
Therefore, improvements from FPTI estimate how much factual degradation is attributable to current-step token perturbations, while any remaining gap indicates the role of prefix-induced drift.

\subsection{Fact-Preserving Attention Intervention}
\label{sec:fpai}

FPAI targets an attention-mediated component of prefix-induced drift.
Unlike FPTI, it does not modify the output-level watermark perturbation \(b_t(y_t;\tilde{y}_{<t})\).
Instead, it restores the model's access to fact-bearing evidence when the watermarked prefix reduces factual attention.

Let \(\mathcal{F}(x)\subseteq\{1,\ldots,|x|\}\) denote the positions of fact-bearing tokens in the input context.
At decoding step \(t\), let \(s_t(j)\) be the pre-softmax attention logit assigned to context token \(j\).
FPAI adds a compensation term \(\lambda_t\) to fact-bearing context positions:
\begin{equation}
\label{eq:fpai}
s_t^{\mathrm{FPAI}}(j)
=
\begin{cases}
s_t(j)+\lambda_t, & j\in \mathcal{F}(x),\\
s_t(j), & j\notin \mathcal{F}(x),
\end{cases}
\qquad
A_t^{\mathrm{FPAI}}
=
\mathrm{softmax}\!\left(s_t^{\mathrm{FPAI}}\right).
\end{equation}
The compensation strength $\lambda_t$ is calibrated to offset the expected attention drop observed under watermarking.
Specifically, we model the decline of the factual attention coefficient \(\alpha_t\) as a function of prefix detectability and generation position:
\begin{equation}
\label{eq:lambda-calibration}
\lambda_t
=
\phi\!\left(Z_{<t},\, r_t\right),
\end{equation}
where \(Z_{<t}\) denotes the watermark score of the generated prefix and \(r_t\) denotes the relative position of the current factual token.
The calibration objective is to restore factual attention toward its clean-decoding level, i.e., \(\alpha_t^{\mathrm{FPAI}}\approx \alpha_t^{\mathrm{clean}}\), without changing the output-level watermark rule.
In practice, \(\phi(\cdot)\) is fitted on held-out calibration data.

\noindent\textbf{Combining two interventions.}
The two interventions are complementary and can be applied jointly.
FPAI restores the evidence-access path by increasing attention to fact-bearing context tokens, while FPTI prevents the subsequent watermark rule from directly perturbing fact-bearing output tokens.
Thus, the combined intervention first improves the contextual basis for factual prediction and then protects the factual token choice itself.
This allows the watermark to remain active on non-factual tokens while reducing both prefix-induced drift and direct token-level perturbations.

\section{Experiments}
\label{sec:empirical}

\subsection{Experimental Setup}
\label{sec:experimental_setup}

\noindent\textbf{Data and task.}
We evaluate watermarking hallucination in a controlled RAG setting~\cite{gao2024retrievalaugmentedgenerationlargelanguage,ni2025trustworthyretrievalaugmentedgeneration}, where each query is paired with a fixed context paragraph containing the target facts.
The evaluation set is constructed from Wikipedia passages associated with
Natural Questions~\cite{kwiatkowski-etal-2019-natural}, using named entity
recognition~\citep{ding-etal-2025-selecting} and rule-based filtering~\citep{tokensregex2014}
to retain verifiable facts.
Queries cover four fact categories: numerical facts, identifiers, temporal facts, and qualitative facts, with 250 samples per category and 1,000 paired generations in total for each experimental condition.

\noindent\textbf{Models and methods.}
Our main experiments use LLaMA 3.1 8B Instruct~\cite{dubey2024llama}.
For cross-model validation, we additionally evaluate Mistral-7B-Instruct-v0.3~\cite{jiang2023mistral7b} and Qwen2.5-14B-Instruct~\cite{qwen2025qwen25technicalreport}.
We incorporate six representative decoding-time watermarking methods: KGW~\cite{pmlr-v202-kirchenbauer23a}, SWEET~\cite{lee-etal-2024-wrote}, DiPmark~\cite{wu2024resilientaccessibledistributionpreservingwatermark}, GumbelSoft~\cite{fu-etal-2024-gumbelsoft}, SynthID-style watermarking~\cite{Dathathri2024ScalableWF}, and Gumbel-Max watermarking~\cite{kuditipudi2024robust}. They cover logit-bias, entropy-aware, distribution-preserving, sampling-based, and tournament-style designs.
We compare the proposed FPTI, FPAI, and their joint application with two representative inference-time hallucination mitigation baselines, DoLa~\cite{chuang2024dola} and DeCoRe~\cite{gema-etal-2025-decore}, under KGW watermarking.
Our interventions differ in that FPTI operates on output logits and watermark perturbations, while FPAI further uses attention logits over the provided context.

\noindent\textbf{Evaluation.}
We report the net factual accuracy decrease \(\Delta_{\mathrm{fact}}\) defined in Equation~\eqref{eq:delta-fact}, with lower values indicating less watermark-induced factual degradation.
Watermark detectability is measured primarily by TPR@1\%FPR.
Because fact-preserving interventions can alter watermark detectability, for each model, watermarking method, and intervention setting, we independently calibrate the watermark strength to restore the same measured TPR@1\%FPR operating point before comparing factual accuracy.
We also report perplexity and decoding speed for trade-off analysis.
Full experimental details are provided in Appendix~\ref{sec:appendix_model_decoding}--\ref{sec:appendix_attention_detectability}.

\begin{table*}[t]
\centering
\small
\setlength{\tabcolsep}{4pt}
\renewcommand{\arraystretch}{0.95}
\caption{
\textbf{Effectiveness of fact-preserving interventions.}
We report the net factual accuracy decrease \(\Delta_{\mathrm{fact}}\) (\%)
across six watermarking methods, three model families, and three matched
TPR@1\%FPR levels.
Lower values indicate less watermark-induced factual degradation.
FPTI limits factual-token perturbations, and FPTI+FPAI further reduces
prefix-induced errors.
}
\label{tab:cross_model_mitigation}

\resizebox{\textwidth}{!}{
\begin{tabular}{l|ccc|ccc|ccc}
\toprule
\multirow{2}{*}{Method / setting}
& \multicolumn{3}{c|}{LLaMA3.1-8B}
& \multicolumn{3}{c|}{Mistral-7B}
& \multicolumn{3}{c}{Qwen2.5-14B} \\
\cmidrule(lr){2-4} \cmidrule(lr){5-7} \cmidrule(lr){8-10}
& TPR=.50 & TPR=.90 & TPR=.98
& TPR=.50 & TPR=.90 & TPR=.98
& TPR=.50 & TPR=.90 & TPR=.98 \\
\midrule

KGW~\cite{pmlr-v202-kirchenbauer23a}
& 2.1\% & 12.9\% & 22.3\%
& 2.0\% & 14.4\% & 23.1\%
& 2.3\% & 13.7\% & 22.7\% \\

\hspace{0.4em}+FPTI
& 1.5\% & 2.4\% & 8.7\%
& 1.5\% & 2.4\% & 8.9\%
& 1.8\% & 2.6\% & 8.9\% \\

\hspace{0.4em}+FPTI+FPAI
& \textbf{0.6\%} & \textbf{0.9\%} & \textbf{6.3\%}
& \textbf{0.8\%} & \textbf{1.2\%} & \textbf{6.3\%}
& \textbf{0.8\%} & \textbf{1.1\%} & \textbf{6.4\%} \\

\midrule

SWEET~\cite{lee-etal-2024-wrote}
& 2.3\% & 15.0\% & 23.1\%
& 2.5\% & 15.6\% & 23.9\%
& 2.9\% & 15.7\% & 24.8\% \\

\hspace{0.4em}+FPTI
& 1.1\% & 1.9\% & 8.6\%
& 1.3\% & 2.0\% & 8.6\%
& 1.1\% & 2.1\% & 9.5\% \\

\hspace{0.4em}+FPTI+FPAI
& \textbf{0.7\%} & \textbf{1.0\%} & \textbf{5.5\%}
& \textbf{0.8\%} & \textbf{1.1\%} & \textbf{5.1\%}
& \textbf{0.7\%} & \textbf{1.3\%} & \textbf{6.1\%} \\

\midrule

DiPmark~\cite{wu2024resilientaccessibledistributionpreservingwatermark}
& 0.7\% & 6.8\% & 8.7\%
& 0.7\% & 7.3\% & 9.2\%
& 0.8\% & 6.5\% & 8.4\% \\

\hspace{0.4em}+FPTI
& 0.4\% & 1.7\% & 4.2\%
& 0.5\% & 1.9\% & 4.6\%
& 0.4\% & 1.6\% & 4.3\% \\

\hspace{0.4em}+FPTI+FPAI
& \textbf{0.2\%} & \textbf{0.4\%} & \textbf{2.6\%}
& \textbf{0.3\%} & \textbf{0.6\%} & \textbf{2.7\%}
& \textbf{0.2\%} & \textbf{0.4\%} & \textbf{2.6\%} \\

\midrule

GumbelSoft~\cite{fu-etal-2024-gumbelsoft}
& 2.0\% & 8.3\% & 18.2\%
& 2.1\% & 8.7\% & 19.2\%
& 2.7\% & 8.1\% & 19.3\% \\

\hspace{0.4em}+FPTI
& 1.3\% & 3.2\% & 7.4\%
& 1.3\% & 3.4\% & 7.7\%
& 1.9\% & 2.8\% & 8.2\% \\

\hspace{0.4em}+FPTI+FPAI
& \textbf{0.7\%} & \textbf{0.8\%} & \textbf{5.6\%}
& \textbf{0.8\%} & \textbf{1.0\%} & \textbf{5.7\%}
& \textbf{0.7\%} & \textbf{0.8\%} & \textbf{5.9\%} \\

\midrule

SynthID-style~\cite{Dathathri2024ScalableWF}
& 0.5\% & 2.7\% & 4.8\%
& 0.6\% & 2.9\% & 4.2\%
& 0.6\% & 2.4\% & 4.5\% \\

\hspace{0.4em}+FPTI
& 0.4\% & 0.9\% & 3.1\%
& 0.4\% & 0.8\% & 3.3\%
& 0.5\% & 0.8\% & 3.3\% \\

\hspace{0.4em}+FPTI+FPAI
& \textbf{0.2\%} & \textbf{0.4\%} & \textbf{2.1\%}
& \textbf{0.3\%} & \textbf{0.5\%} & \textbf{2.3\%}
& \textbf{0.3\%} & \textbf{0.4\%} & \textbf{2.7\%} \\

\midrule

Gumbel-Max~\cite{kuditipudi2024robust}
& 0.6\% & 3.7\% & 5.5\%
& 0.7\% & 3.6\% & 5.7\%
& 0.6\% & 4.1\% & 5.2\% \\

\hspace{0.4em}+FPTI
& 0.4\% & 1.1\% & 3.5\%
& 0.5\% & 1.0\% & 3.7\%
& 0.5\% & 1.3\% & 3.7\% \\

\hspace{0.4em}+FPTI+FPAI
& \textbf{0.2\%} & \textbf{0.5\%} & \textbf{2.4\%}
& \textbf{0.3\%} & \textbf{0.6\%} & \textbf{2.6\%}
& \textbf{0.2\%} & \textbf{0.7\%} & \textbf{2.5\%} \\

\bottomrule
\end{tabular}
}
\end{table*}

\subsection{Performance of Proposed Fact-Preserving Interventions}
\label{sec:mitigation_results}

\noindent\textbf{Main intervention results.}
We first evaluate whether the proposed interventions reduce watermarking hallucination across watermarking methods, model families, and matched detectability levels.
Specifically, we compare standard watermarking with its FPTI and FPTI+FPAI variants, where FPTI removes current-step perturbations on fact-bearing tokens and FPAI further compensates for prefix-induced loss of factual attention.
The results in \Cref{tab:cross_model_mitigation} show a consistent reduction in \(\Delta_{\mathrm{fact}}\) from standard watermarking to FPTI and further to FPTI+FPAI.
For example, under KGW on LLaMA3.1-8B at \(\mathrm{TPR@1\%FPR}=0.90\), \(\Delta_{\mathrm{fact}}\) drops from \(12.9\%\) to \(2.4\%\) with FPTI and further to \(0.9\%\) with FPTI+FPAI.
Similar reductions appear across Mistral-7B and Qwen2.5-14B, indicating that the effect is not tied to one model family.
Nevertheless, residual errors remain more visible under very high detectability, suggesting that strong watermarking can push the prefix away from the evidence before later factual tokens are locally protected or attention-compensated.

The table also compares the intrinsic factuality cost of different watermarking rules under the same detectability target.
The results show that watermarking methods differ substantially in how much factual degradation they introduce.
For example, on LLaMA3.1-8B at \(\mathrm{TPR@1\%FPR}=0.90\), SynthID-style watermarking has a much lower \(\Delta_{\mathrm{fact}}\) than KGW or DiPmark before intervention.
This suggests that more controlled or distribution-aware sampling rules may be less prone to factual drift than stronger logit-bias-style perturbations, although all methods still benefit from FPTI+FPAI.

\begin{figure}[t]
\centering
\includegraphics[width=0.86\linewidth]{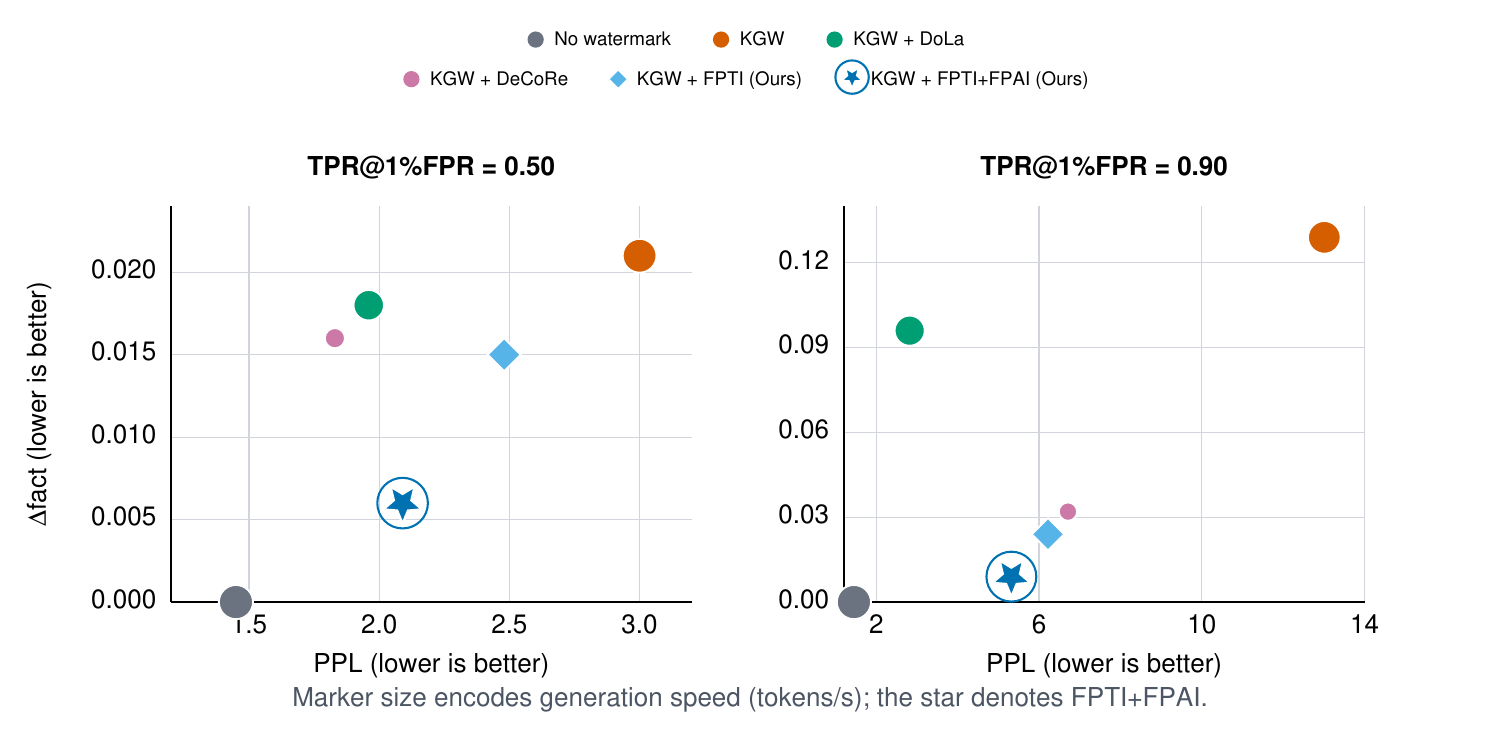}

\caption{
\textbf{Comparing trade-offs under matched detectability.}
Lower-left (lower PPL and lower \(\Delta_{\mathrm{fact}}\)) is better.
Marker size denotes decoding speed, with larger circles indicating higher speed.
}

\label{fig:tradeoff_factuality_fluency}
\end{figure}

\noindent\textbf{Comparison with general hallucination mitigation.}
We further compare FPTI and FPAI with DoLa~\cite{chuang2024dola} and DeCoRe~\cite{gema-etal-2025-decore} under matched KGW detectability.
Table~\ref{tab:factuality_fluency_efficiency} reports the resulting \(\Delta_{\mathrm{fact}}\), perplexity, and decoding-speed trade-offs, and Figure~\ref{fig:tradeoff_factuality_fluency} visualizes the same trade-off under matched detectability.
All mitigation methods improve over standard KGW by reducing both \(\Delta_{\mathrm{fact}}\) and PPL, but their trade-offs reflect different mechanisms.
DoLa improves fluency: at \(\mathrm{TPR@1\%FPR}=0.90\), it reduces PPL from \(12.98\) to \(2.84\), but \(\Delta_{\mathrm{fact}}\) remains relatively high at \(9.6\%\).
This suggests that layer-contrast decoding can make generations more fluent without directly correcting watermark-induced factual shifts.
DeCoRe is more effective for factuality, lowering \(\Delta_{\mathrm{fact}}\) to \(3.2\%\), because retrieval-head contrast is better aligned with context grounding; however, it still does not remove current-step watermark perturbations on factual tokens and incurs a larger decoding cost.

By contrast, FPTI+FPAI directly targets the two error sources identified in our mechanism analysis.
FPTI removes current-step watermark perturbations on fact-bearing tokens, while FPAI compensates for prefix-induced loss of factual attention.
As a result, FPTI alone reduces \(\Delta_{\mathrm{fact}}\) to \(2.4\%\), outperforming DeCoRe in factuality with almost no speed loss; adding FPAI further reduces \(\Delta_{\mathrm{fact}}\) to \(0.9\%\), achieving a \(93.0\%\) reduction over KGW while maintaining KGW-like speed.
These results indicate that watermarking hallucination is not merely a generic hallucination or fluency problem; it is best mitigated by interventions that explicitly account for how watermarking perturbs factual token selection and evidence use.

\begin{table*}[t]
\centering
\small
\setlength{\tabcolsep}{4pt}

\caption{
\textbf{Comparing proposed interventions with hallucination mitigation baselines.}
We compare our interventions with the hallucination mitigation baselines
DoLa~\citep{chuang2024dola} and DeCoRe~\citep{gema-etal-2025-decore}
under KGW watermarking~\cite{pmlr-v202-kirchenbauer23a}.
\(\Delta_{\mathrm{fact}}\) measures the net factual accuracy decrease,
and Improv. denotes its relative reduction over KGW.
PPL measures fluency, and Speed is measured in generated tokens per second.
}

\label{tab:factuality_fluency_efficiency}
\resizebox{\textwidth}{!}{
\begin{tabular}{l|cccc|cccc}
\toprule
\multirow{2}{*}{Setting}
& \multicolumn{4}{c|}{\textbf{TPR@1\%FPR = 0.50}}
& \multicolumn{4}{c}{\textbf{TPR@1\%FPR = 0.90}} \\
\cmidrule(lr){2-5} \cmidrule(lr){6-9}
& $\Delta_{\mathrm{fact}}$ $\downarrow$ & Improv. $\uparrow$ & PPL $\downarrow$ & Speed $\uparrow$
& $\Delta_{\mathrm{fact}}$ $\downarrow$ & Improv. $\uparrow$ & PPL $\downarrow$ & Speed $\uparrow$ \\
\midrule

No watermark
& 0.0\% & -- & 1.45 & 1.52
& 0.0\% & -- & 1.45 & 1.52 \\

KGW~\citep{pmlr-v202-kirchenbauer23a} (baseline)
& 2.1\% & 0.0\% & 3.01 & 1.49
& 12.9\% & 0.0\% & 12.98 & 1.48 \\

KGW~\citep{pmlr-v202-kirchenbauer23a} + DoLa~\citep{chuang2024dola}
& 1.8\% & 14.3\% & 1.98 & 1.41
& 9.6\% & 25.6\% & \textbf{2.84} & 1.40 \\

KGW~\citep{pmlr-v202-kirchenbauer23a} + DeCoRe~\citep{gema-etal-2025-decore}
& 1.6\% & 23.8\% & \textbf{1.86} & 1.14
& 3.2\% & 75.2\% & 6.74 & 1.10 \\

KGW~\citep{pmlr-v202-kirchenbauer23a} + FPTI (Ours)
& 1.5\% & 28.6\% & 2.47 & \textbf{1.48}
& 2.4\% & 81.4\% & 6.20 & \textbf{1.47} \\

KGW~\citep{pmlr-v202-kirchenbauer23a} + FPTI+FPAI (Ours)
& \textbf{0.6\%} & \textbf{71.4\%} & 2.07 & 1.46
& \textbf{0.9\%} & \textbf{93.0\%} & 5.35 & 1.46 \\

\bottomrule
\end{tabular}
}
\end{table*}

\vspace{-2mm}

\begin{wrapfigure}[16]{r}{0.40\linewidth}
\vspace{-6mm}
\centering
\includegraphics[width=\linewidth]{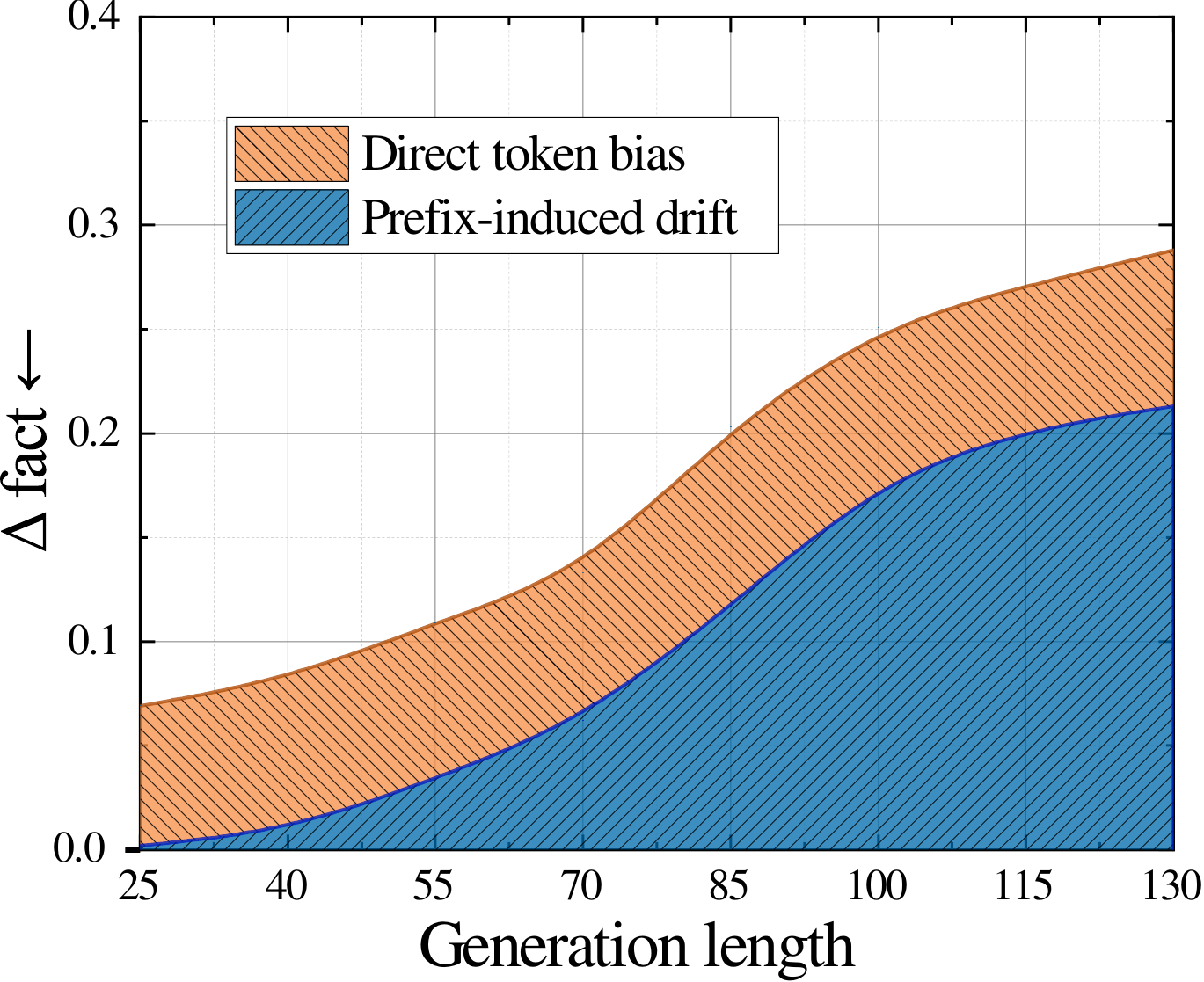}
\caption{
Position-dependent residual error under KGW. Blue: residual \(\Delta_{\mathrm{fact}}\) after FPTI; orange: reduction in \(\Delta_{\mathrm{fact}}\) by FPTI vs. standard watermarking.
}
\label{fig:token_attention}
\end{wrapfigure}

\subsection{Position-Dependent Residual Error}
\label{sec:position_dependent_residual}

We use FPTI to isolate current-step token perturbations from prefix-accumulated errors.
For each target query, we vary its prompt position while fixing the evidence block, target fact, template, sampling parameters, and watermark detectability.
We then measure \(\Delta_{\mathrm{fact}}\) with and without FPTI, grouped by the generation position of the factual token.

\Cref{fig:token_attention} visualizes the resulting decomposition by generation position.
The gap between standard watermarking and FPTI estimates the current-step token component removed by FPTI, which remains relatively stable across positions.
In contrast, the residual after FPTI grows substantially with generation depth and becomes the dominant error source after generating 70 tokens.
This further supports the watermark avalanche effect: early watermark-conditioned choices gradually move the prefix away from the evidence-supported trajectory, so later factual decisions remain vulnerable even when their own watermark perturbations are removed.

This final analysis shows the boundary of local repair.
Fact-preserving watermarking cannot rely only on protecting the current factual token; it must also limit prefix-induced drift before factual decisions are made, especially in fact-critical applications where small substitutions in dates, identifiers, or numerical values can be consequential.

\section{Discussion and Conclusion}
\label{sec:discussion_conclusion}

This paper studies factual degradation caused by watermarking itself.
We identify \emph{watermarking hallucination} as a failure mode distinct from ordinary hallucination and surface-quality loss: watermarked text can remain fluent and detectable while changing a context-supported fact.
Using the net factual accuracy decrease \(\Delta_{\mathrm{fact}}\), we show that this effect appears across watermarking methods, fact types, and model families.
We attribute it to two coupled sources: current-step token perturbations and prefix-induced attention interference.
FPTI and FPAI validate this mechanism by protecting fact-bearing output tokens and restoring attention to factual context, substantially reducing watermark-induced errors.

Our study also reveals the boundary of local repair.
Under strong watermarking, residual errors remain.
Early watermark-conditioned choices can push the prefix away from the supporting evidence before later factual decisions occur.
This suggests that factuality should be treated as a watermark design constraint, not only a post-hoc metric.
Future watermarks should place their signal where it least interferes with grounding, avoiding numerical values, identifiers, dates, citations, and prefix positions that affect later factual attention.
Overall, watermarking should not be assumed harmless in fact-critical applications: factual reliability must be evaluated alongside detectability, robustness, and output quality.

\renewcommand{\acksection}{\section*{Acknowledgments}}
\begin{ack}
We thank the anonymous reviewers for their helpful suggestions. This work is supported by the National Key Research and Development Program of China under (Grant No.2023YFB2703900).
\end{ack}

\clearpage
\appendix

\section{Related Works}
The growing deployment of AI systems has brought increasing
attention to their security, including model ownership protection
and robustness against adversarial threats~\cite{zhang2025mer,hu2024veridip,zhang2026on,11422282,11612913}.
As AI-generated content becomes more widespread, tracing its
provenance is another important concern.
Text watermarking offers a means of identifying model-generated
content, but its effects on factual reliability also require
careful evaluation. We review the literature on LLM watermarking, factuality risks
in watermarked text, context-grounded generation and evaluation,
and decoding-time methods for improving factuality.

\noindent\textbf{LLM watermarking.}
LLM watermarking methods embed detectable signals into generated text by modifying decoding, constraining sampling, or preserving distributional properties while adding a hidden statistical structure~\cite{pmlr-v202-kirchenbauer23a,kuditipudi2024robust,Dathathri2024ScalableWF,liu2024surveytextwatermarking,yang2025watermarksurvey}. 
KGW-style methods bias token sets directly, while methods such as DiPmark, SWEET, GumbelSoft, token-specific watermarking, and distribution-adaptive watermarking seek better quality and robustness trade-offs~\cite{wu2024resilientaccessibledistributionpreservingwatermark,lee-etal-2024-wrote,fu-etal-2024-gumbelsoft,10.5555/3692070.3692903,he2025dawa}. 
Robustness studies examine paraphrase attacks, watermark removal, watermark stealing, and adaptive evasion attacks~\cite{rastogi-pruthi-2024-revisiting,wu-chandrasekaran-2024-bypassing,10.5555/3692070.3692977,pmlr-v267-diaa25a}. 
These approaches are usually evaluated through detectability, false positive rate, robustness to edits, and surface quality. 
Our work studies a different axis: whether the same decoding changes also harm factual reliability.

\noindent\textbf{Factuality risks in watermarked text generation.}
\Citet{hastuti-etal-2025-factuality} show that coherence-oriented evaluation is insufficient for medical watermarked generation and introduce a GPT-based judging workflow for factuality-aware assessment. 
Their results suggest that factual degradation can be more severe than coherence loss, especially when watermarking shifts entity-level or domain-specific content. 
Our work builds on this observation but differs in scope and goal. 
First, we study watermarking hallucination under a paired context-grounded setting, where the relevant evidence is explicitly available and the unwatermarked model can answer correctly. 
Second, rather than measuring factuality as an absolute output property, we use \(\Delta_{\mathrm{fact}}\) to measure the net factual accuracy decrease between paired unwatermarked and watermarked generations. 
Third, we go beyond evaluation by attributing the degradation to current-step token perturbations and prefix-induced attention interference, and by introducing fact-preserving interventions that target these mechanisms.

\noindent\textbf{Context-grounded factuality and RAG evaluation.}
Retrieval-augmented generation is widely used to reduce hallucination by conditioning generation on external evidence~\cite{gao2024retrievalaugmentedgenerationlargelanguage,ni2025trustworthyretrievalaugmentedgeneration}. 
However, the presence of relevant context does not by itself guarantee faithful output: models may ignore the supplied evidence, rely on parametric priors, misread retrieved passages, or drift away from the context during long-form generation~\cite{tao-etal-2024-context,xu-etal-2024-knowledge-conflicts}. 
Recent work further studies faithful RAG through internal model signals, for example using sparse autoencoders to identify features associated with source-faithful and unfaithful generation~\cite{xiong2026toward}. 
Factuality evaluation methods therefore increasingly emphasize evidence-grounded verification rather than surface fluency alone. 
FActScore decomposes long-form answers into atomic facts and checks whether they are supported~\cite{min-etal-2023-factscore}, while SAFE uses structured search and verification to evaluate long-form factuality~\cite{wei2024longform}. 
Our evaluation follows the same evidence-grounded principle but applies it to a different causal question: given the same model, prompt, retrieved context, and sampling configuration, does watermarking itself introduce factual errors? 
This paired design separates watermark-induced factual degradation from ordinary model errors or retrieval failures.

\noindent\textbf{Hallucination and decoding-time factuality.}
Hallucination research identifies failures caused by missing knowledge, knowledge conflict, weak grounding, uncertainty, and inference-time dynamics~\cite{10.1145/3703155,xu-etal-2024-knowledge-conflicts,tao-etal-2024-context,farquhar2024semanticentropy}. 
A complementary line of work studies the knowledge boundary of LLMs, where factual failures arise because models cannot reliably memorize, access, or apply certain types of knowledge~\cite{li-etal-2025-knowledge-boundary}. 
Decoding interventions can improve or degrade factuality, as shown by methods such as ITI and DoLa~\cite{NEURIPS2023_81b83900,chuang2024dola}. 
These methods treat hallucination as a general generation failure and aim to make the model more faithful. 
In contrast, \wh is induced by a source-attribution mechanism inserted for detectability: watermark-conditioned token choices can alter the generated prefix and later weaken attention to factual evidence. 
This makes \wh a distinct inference-time failure mode: the model may be capable of using the evidence under clean decoding, but the watermark shifts the generation trajectory away from that evidence. 
Our FPTI and FPAI interventions are therefore not general hallucination mitigators; they are watermark-aware probes and repairs designed to separate and reduce the two mechanisms responsible for watermark-induced factual errors.

\section{Experimental Details}
\label{sec:appendix_exp}

\subsection{Model and Decoding Setup}
\label{sec:appendix_model_decoding}

We use instruction-tuned open-weight models throughout the experiments. 
LLaMA 3.1 8B Instruct~\cite{dubey2024llama} is used as the main model for the mechanism analysis and the primary watermarking evaluation. 
To test whether the observed trends generalize across model families, we additionally evaluate Mistral-7B-Instruct-v0.3~\cite{jiang2023mistral7b} and Qwen2.5-14B-Instruct~\cite{qwen2025qwen25technicalreport} in the cross-model intervention study.

For watermarking, we evaluate six representative decoding-time watermarking paradigms: KGW~\cite{pmlr-v202-kirchenbauer23a}, SWEET~\cite{lee-etal-2024-wrote}, DiPmark~\cite{wu2024resilientaccessibledistributionpreservingwatermark}, GumbelSoft~\cite{fu-etal-2024-gumbelsoft}, SynthID-style watermarking~\cite{Dathathri2024ScalableWF}, and Gumbel-Max-style watermarking~\cite{kuditipudi2024robust}. 
Our implementations follow their corresponding published mechanisms and cover logit-bias, entropy-aware, distribution-preserving, sampling-path, and tournament-style designs.
For mitigation, we evaluate the proposed FPTI and FPAI interventions. 
In the comparison with general hallucination mitigation methods, we further include DoLa~\cite{chuang2024dola} and DeCoRe~\cite{gema-etal-2025-decore} under KGW watermarking.

For each condition, the unwatermarked and watermarked outputs are generated with the same prompt, context, sampling parameters, and random-seed schedule. 
This pairing is important because the goal is not to estimate general model factuality, but to isolate the factual change caused by adding a watermark. 
When comparing watermarking methods or interventions, we align settings by watermark detectability whenever possible rather than by nominal strength, since different watermarking rules can produce different detector performance under the same strength parameter.

Licenses and existing assets.
We use publicly available datasets, open-weight models, and open-source software packages. The RAG contexts are constructed from Wikipedia passages associated with Natural Questions~\cite{kwiatkowski-etal-2019-natural}, whose official repository is released under Apache-2.0. LLaMA 3.1 8B Instruct is used under the Meta Llama 3.1 Community License. Mistral-7B-Instruct-v0.3 and Qwen2.5-14B-Instruct are released under Apache-2.0. For watermarking and decoding baselines, we cite the corresponding papers and use either publicly available implementations when their licenses permit research use or our own implementations based on the published descriptions. 
We do not redistribute model weights or original dataset contents. 
The implementation includes paired-generation scripts, intervention-generation scripts, structured annotation utilities, and toy examples for smoke testing; full experiments require evaluation samples following the JSONL schema described in Appendix~\ref{sec:appendix_code_data}.

Table~\ref{tab:decoding_config} summarizes the decoding configuration used for unwatermarked, watermarked, and intervention-enhanced generations.

\begin{table}[h]
\centering
\small
\caption{\textbf{Decoding configuration.} The same decoding parameters are used for unwatermarked, watermarked, and intervention-enhanced generations unless an experiment explicitly varies watermark strength or detectability.}
\label{tab:decoding_config}
\begin{tabular}{lc|lc}
\toprule
\textbf{Parameter} & \textbf{Value}
& \textbf{Parameter} & \textbf{Value} \\
\midrule
Temperature & 0.7
& Top-\(p\) & 0.9 \\

Top-\(k\) & 40
& Repetition penalty & 1.1 \\

No-repeat n-gram size & 3
& Maximum input context & 300 tokens \\

Maximum new tokens & 300
& Sampling & Enabled \\
\bottomrule
\end{tabular}
\end{table}

\subsection{RAG Evaluation Setup}
\label{sec:appendix_rag_setup}

We construct a controlled retrieval-augmented generation (RAG) setting, 
where each query is paired with a fixed context paragraph containing multiple verifiable facts.

The evaluation set is built from Wikipedia passages associated with Natural Questions. 
We apply named entity recognition (NER) together with rule-based filtering to identify candidate contexts, 
and retain those that contain at least three facts of the same type. 
The resulting contexts span diverse topics (e.g., legal, medical, and financial domains) 
and cover four categories of factual information.

Each evaluation sample is designed to target a specific category of facts. 
For each query, the model is required to extract and answer one or more target
facts of the same type from the provided context within a single response.

We evaluate four categories of factual information, covering different types of reasoning and representation:

\textbf{Numerical facts} refer to quantitative values (e.g., ``78 mg/dL''), 
which require precise numerical correctness and sensitivity to units.

\textbf{Identifier facts} include named entities, codes, or symbolic references 
(e.g., ``AAPL'', ``Doe v. Smith''), where correctness depends on exact matching.

\textbf{Temporal facts} correspond to dates and time expressions (e.g., ``May 12''), 
which must be chronologically accurate.

\textbf{Qualitative facts} represent descriptive or evaluative information 
(e.g., ``relatively minor''), which involves comparative or contextual interpretation.

For each fact type, we construct 250 evaluation samples, resulting in 1,000 paired generations for each experimental condition and a balanced test set across factual categories.

\paragraph{Data source and privacy.}
Our evaluation data are constructed from publicly available Wikipedia passages associated with Natural Questions. 
We do not collect new user data or private information. 
The dataset is used only for controlled factuality evaluation, and each item is derived from a public context paragraph paired with a synthetic or filtered query. 
Although the contexts cover diverse topics, the benchmark is not intended to represent all fact-critical domains, such as clinical, legal, or financial deployment settings.

\subsection{Net Factual Accuracy Decrease Evaluation Protocol}
\label{sec:appendix_delta_fact_protocol}

We evaluate watermark-induced factual degradation using the paired metric defined in
\Cref{eq:delta-fact}. For each evaluation item, we compare the unwatermarked output
with the corresponding watermarked output under the same context, query,
sampling configuration, and random-seed schedule. We record both
correct-to-incorrect and incorrect-to-correct transitions and compute
\(\Delta_{\mathrm{fact}}\) as their difference over the full evaluation set.

The factuality evaluator checks whether the answer is supported by the provided
context. For numerical, temporal, and identifier facts, we use exact or
normalized matching whenever possible, allowing harmless surface variations
such as equivalent date formats or unit spellings. For qualitative facts, we
use structured annotation followed by manual verification on sampled outputs.
A watermarked answer is marked incorrect if it omits a required target fact,
changes a context-supported value, or introduces a contradictory claim.

This paired protocol differs from absolute factuality judging. It does not ask
whether a model is generally factual across open-ended tasks. Instead, it measures
the change in factual accuracy caused by watermarking while accounting for
transitions in both directions.

We do not rely on an LLM judge for the main reported factuality labels. 
Rule-based decisions are manually audited on sampled outputs to check whether normalization and matching rules correctly capture factual support.

\subsection{Attention Measurement and Detectability}
\label{sec:appendix_attention_detectability}

To compute the factual attention coefficient \(\alpha_t\), we first identify fact-bearing context tokens associated with the target facts. We then aggregate attention mass assigned to those tokens across layers and heads at each decoding step. The resulting \(\alpha_t\) is averaged over generated samples within the same watermarking method and detectability regime. This measurement is not used as a standalone explanation of model reasoning; it is used as an observable signal whose causal relevance is tested through FPAI in the main text.

Detectability is aligned by grouping runs into low-, mid-, and high-Z regimes when the detector reports a Z-score. This avoids comparing methods at the same nominal hyperparameter when the actual watermark signal differs. For methods without a directly comparable Z-score, appendix plots use the method's native strength axis.

\subsection{Representative Intervention Cases}

\Cref{tab:case_success} shows a case where the combined intervention restores the target facts. The standard watermark produces plausible but wrong numerical values. FPTI+FPAI recovers the original values because the early part of the answer still follows the evidence closely enough for attention restoration to work.

\begin{table}[h]
\centering
\footnotesize
\setlength{\tabcolsep}{3pt}
\renewcommand{\arraystretch}{1.05}
\caption{\textbf{Successful recovery under FPTI+FPAI.} The standard watermark produces plausible but incorrect numerical values, while FPTI+FPAI restores the context-supported answer.}
\label{tab:case_success}
\begin{tabularx}{\linewidth}{p{0.20\linewidth}|X|c}
\toprule
\textbf{Setting} & \textbf{Output} & \textbf{Factual?} \\
\midrule
No watermark
& Fasting glucose range: 145--162 mg/dL; postprandial peak: 210 mg/dL; nocturnal low: 78 mg/dL.
& Yes \\
Watermark
& Fasting glucose ranged 145--160 mg/dL; peak value 200--210 mg/dL; nocturnal low around 73 mg/dL.
& No \\
FPTI+FPAI
& Fasting sugar level: 145--162 mg/dL; peak value: 210 mg/dL; nocturnal low: 78 mg/dL.
& Yes \\
\bottomrule
\end{tabularx}
\end{table}

\Cref{tab:case_failure} shows the limit of local repair. Once the watermarked prefix becomes malformed, restoring factual tokens and attention can recover part of the answer but cannot reconstruct the full unwatermarked trajectory.

\begin{table}[h]
\centering
\footnotesize
\setlength{\tabcolsep}{3pt}
\renewcommand{\arraystretch}{1.05}
\caption{\textbf{Incomplete recovery under FPTI+FPAI.} Severe prefix drift leaves residual factual errors after both interventions.}
\label{tab:case_failure}
\begin{tabularx}{\linewidth}{p{0.20\linewidth}|X|c}
\toprule
\textbf{Setting} & \textbf{Output} & \textbf{Factual?} \\
\midrule
No watermark
& Fasting glucose range: 145--162 mg/dL; postprandial peak: 210 mg/dL; nocturnal low: 78 mg/dL.
& Yes \\
Watermark
& Fasting glucrose range for this particular patient span from approximately between 145--155 to 148 milligrm/dltrange.
& No \\
FPTI+FPAI
& Fasting sugar readings' range: 145--162 mg/dL. Specific value was not listed.
& No \\
\bottomrule
\end{tabularx}
\end{table}

\subsection{Compute Resources}
All experiments are performed on a cluster equipped with NVIDIA GeForce RTX 4090 GPU. 
A single strength-setting run took approximately 70,000--110,000 seconds. 
For each watermarking method, the cross-model mitigation sweep over 8 strengths, 3 settings, and 3 models required approximately 560--880 GPU hours. 
The total compute varies depending on the number of watermarking methods included in each reported experiment; preliminary debugging and failed runs are not included in this estimate.

\subsection{Code and Data Availability}
\label{sec:appendix_code_data}

The code is available at \url{https://github.com/Haocheng-NJUST/WM-Hallucination}. Our implementation covers paired RAG generation, watermark detectability logging, FPTI/FPAI intervention generation, and structured annotation aggregation.
It reads evaluation items from a JSONL schema containing the context, query, fact type, and target facts.
Generation scripts do not infer final factuality labels from raw text; final labels are loaded from structured annotation files, with qualitative cases manually verified as described above. 
The implementation includes toy examples for smoke testing the pipeline.
We do not redistribute model weights or original dataset contents; users should obtain open-weight models and public datasets from their official sources.

\section{Additional Analysis}

\subsection{Additional Watermark Families}
\label{sec:appendix_additional_watermarks}

We further evaluate Unbiased Watermark with \(\delta\)-reweighting~\cite{hu2024unbiased}, TextSeal~\cite{sander2026textseallocalizedllmwatermark}, and MorphMark~\cite{wang2025morphmarkflexibleadaptivewatermarking} to cover additional watermark designs.
For each method, we evaluate 1,000 paired generations covering four fact categories, with 250 examples per category.
The watermark strength is calibrated to a measured \(\mathrm{TPR@1\%FPR}=0.90\).

\begin{table}[t]
\centering
\small
\setlength{\tabcolsep}{4pt}
\caption{
\textbf{Factual transitions for additional watermark families.}
Correct-to-incorrect and incorrect-to-correct transitions use the full set of 1,000 paired generations as their denominator.
The net factual accuracy decrease is their difference.
}
\label{tab:additional_watermark_families}
\resizebox{\linewidth}{!}{
\begin{tabular}{lccc}
\toprule
Method
& Correct$\rightarrow$Incorrect
& Incorrect$\rightarrow$Correct
& $\Delta_{\mathrm{fact}}$ [95\% CI] \\
\midrule

Unbiased Watermark (\(\delta\)-reweighting)
& 7.1\%
& 3.5\%
& 3.6\% [3.1\%, 3.9\%] \\

TextSeal
& 7.7\%
& 3.8\%
& 3.9\% [1.8\%, 5.8\%] \\

MorphMark
& 16.4\%
& 4.3\%
& 12.1\% [11.5\%, 13.7\%] \\

\bottomrule
\end{tabular}
}
\end{table}

All three methods produce transitions in both directions, although the incorrect-to-correct transitions do not fully offset the correct-to-incorrect transitions.
Unbiased Watermark and TextSeal yield relatively small net factual accuracy decreases of 3.6 and 3.9 percentage points, respectively, whereas MorphMark produces a substantially larger decrease of 12.1 points.
Together with the results in \Cref{fig:WHR_ALL,tab:cross_model_mitigation}, these findings show that the magnitude of watermark-induced factual degradation depends on the watermarking design.

\begin{table*}[t]
\centering
\caption{
\textbf{Fluent does not imply factual.}
The watermarked output by SynthID~\cite{Dathathri2024ScalableWF} has \emph{lower perplexity} but introduces an entity-level factual error, while the non-watermarked output remains correct.
In our evaluation under the standard SynthID-style setting, such entity-level watermarking hallucinations occur in 1.7\% of cases.
}
\vspace{-2mm}
\resizebox{\textwidth}{!}{%
\begin{tabular}{p{5.2cm}|p{5.2cm}|p{5.2cm}}
\toprule
\multicolumn{3}{c}{
{\small \textbf{Question:} What case sets the key precedent for liability standards in the Ninth Circuit?}
}
\\
\midrule
\textbf{Context (RAG)} & \textbf{No Watermark} & \textbf{With Watermark (SynthID)} \\
\midrule
\textbf{...}

\textbf{Legal context}

\textit{Precedent:} \textbf{Doe v. Smith}

\textit{Citation:} 2023 WL 183472 (9th Cir.)

\textit{Summary:} Establishes supervisory liability standards in the Ninth Circuit.

\textbf{...}
&
The controlling precedent is \textcolor{green!50!black}{\textbf{Doe v. Smith}}, 2023 WL 183472 (9th Cir.), which establishes the framework for supervisory liability.

\vspace{4pt}
{\footnotesize \textcolor{green!50!black}{\textit{Factually correct}}}

\vspace{6pt}
\textbf{PPL = 4.17}

&
The controlling authority is \textcolor{red}{\textbf{Doe v. Smithson}}, 2023 WL 183472 (9th Cir.).

\vspace{15pt}
{\footnotesize \textcolor{red}{\textit{Entity-level factual error}}}

\vspace{6pt}
\textbf{PPL = 3.31} \hspace{4pt} \textbf{(\textcolor{red}{lower})}
\\
\bottomrule
\end{tabular}%
}
\label{tab:qualitative_fluent_not_factual}
\end{table*}

\subsection{Fluency Metrics and Factual Errors}
\label{sec:appendix_fluency_factuality}

\begin{figure}[h]
\centering
\includegraphics[width=0.68\linewidth]{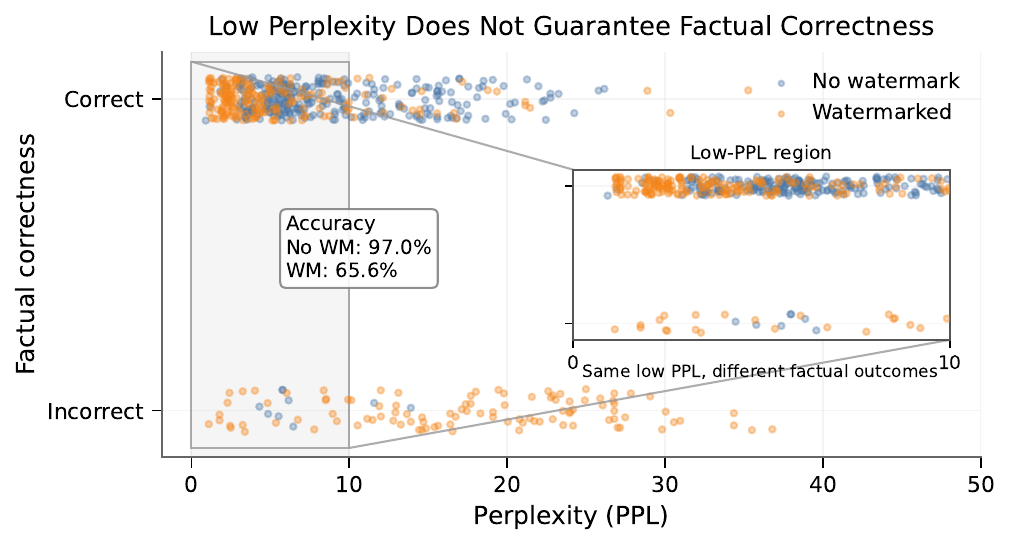}
\caption{\textbf{Token-level perplexity versus factual correctness.} Many incorrect factual tokens appear in low-perplexity regions, showing why fluency-oriented metrics can miss watermark-induced factual errors.}
\label{fig:ppl_vs_factual}
\end{figure}

We examine whether token-level perplexity separates factually correct and incorrect tokens. For each generated token with a factual correctness label, we compute its perplexity under the base model. \Cref{fig:ppl_vs_factual} shows that incorrect factual tokens are not restricted to high-perplexity regions. Correct and incorrect tokens overlap substantially, including in the low-perplexity region highlighted by the inset.

This analysis explains the example in \Cref{tab:qualitative_fluent_not_factual}. A wrong token can be easy for the language model to generate if it fits the local sentence pattern. Factuality depends on whether the token matches the supplied evidence, not only on whether the token is likely under the model. Watermark evaluation therefore needs paired factuality measures such as \(\Delta_{\mathrm{fact}}\) in addition to detectability and fluency checks.

\subsection{Entropy Does Not Identify Factual Tokens}
\label{sec:appendix_entropy_factual_tokens}

SWEET avoids watermarking low-entropy positions, which makes it a natural baseline for reducing quality degradation. 
However, low entropy is not the same as factual importance. 
A token can be factual even when the model assigns high uncertainty to it, especially for numerical values, identifiers, and dates where multiple surface forms may be locally plausible but only one is supported by the context.

\begin{figure}[t]
\centering
\includegraphics[width=0.68\linewidth]{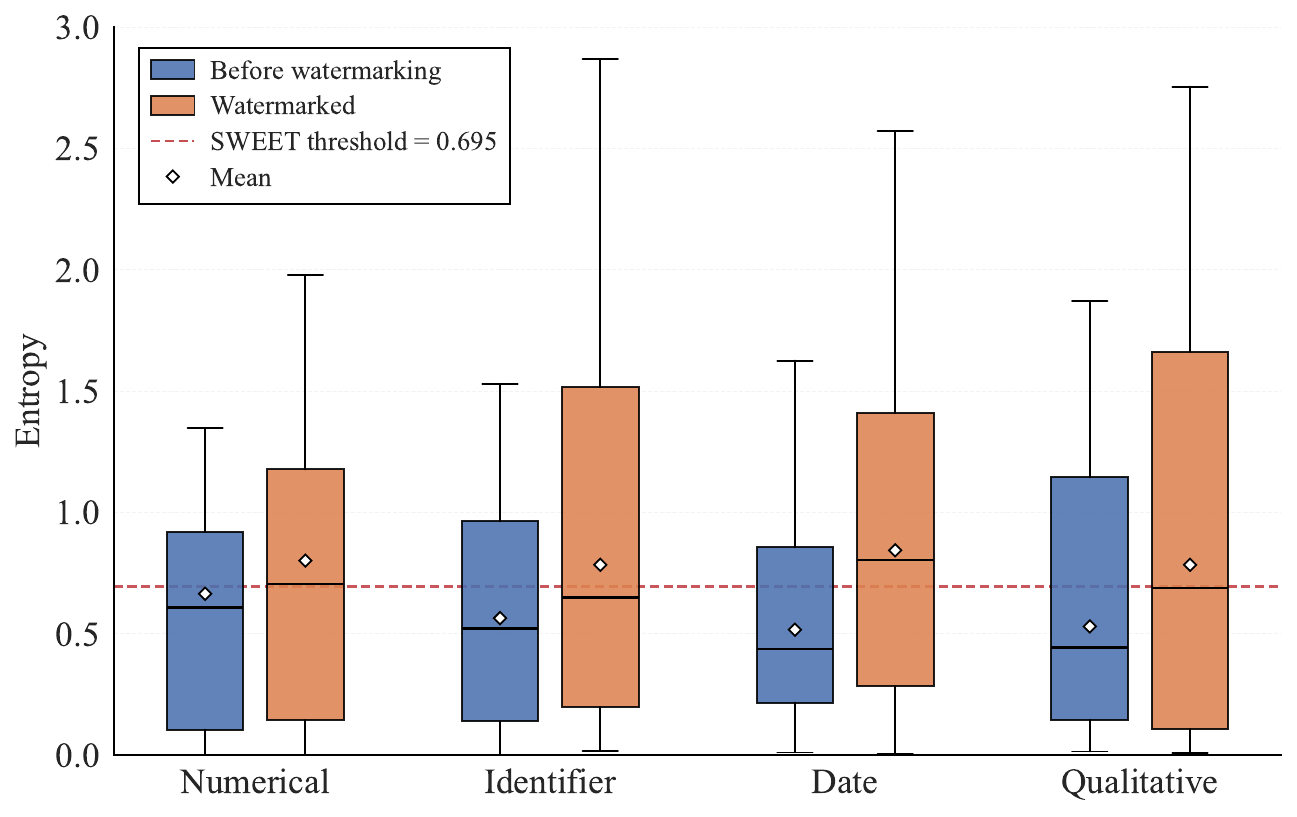}
\caption{
\textbf{Entropy of factual tokens before and after watermarking.}
We report the mean token-level entropy for factual tokens across four fact types. 
The dashed line marks the SWEET entropy threshold (\(0.695\)). 
Many factual tokens lie above this threshold, indicating that entropy-based protection does not reliably cover all fact-bearing tokens.
}
\label{fig:entropy_factual_tokens}
\end{figure}

As shown in \Cref{fig:entropy_factual_tokens}, factual tokens are not confined to the low-entropy region protected by SWEET. 
This explains why entropy-aware watermarking can still produce watermarking hallucination: protecting low-entropy positions preserves some high-confidence generation choices, but many factual decisions occur at higher-entropy positions where the watermark remains active. 
Consequently, SWEET reduces perturbations for a subset of tokens that are important for generation quality, but it does not provide explicit protection for all fact-bearing tokens. 
This suggests that entropy-aware protection can reduce some generation-quality perturbations, but it does not by itself guarantee fact preservation.

\section{Broader Impact}
\label{app:broadimpact}
This work can contribute to a safer and more responsible deployment of LLM watermarking by revealing a factual reliability risk that is not captured by standard detectability, robustness, or fluency metrics. Our study does not argue against watermarking. 
Instead, it shows that watermark deployment in fact-critical applications should be accompanied by factuality evaluation and domain-specific validation. 
By identifying watermark-induced hallucination and proposing fact-preserving interventions, our study can help model providers, policymakers, and application developers evaluate watermarking systems more comprehensively before deploying them in fact-critical settings. 
The proposed analysis may also encourage future watermark designs to preserve factual reliability while maintaining provenance and accountability benefits.
This work may also have unintended negative impacts. 
The finding that watermarking can induce factual errors could be misinterpreted as evidence that watermarking is broadly unreliable, potentially weakening trust in legitimate provenance mechanisms. 
In addition, our failure-mode analysis may help adversarial users identify settings where watermarked generation is more vulnerable to factual errors, such as strong watermarking or long generations. 
These risks highlight the need to interpret our findings as a call for more careful watermark evaluation and deployment, rather than as a rejection of watermarking itself. 

\section{Limitations}
\label{sec:limitations_societal}
Although our interventions reduce watermark-induced hallucinations, they do not completely eliminate factual degradation under strong watermarking or long generations, because a realized watermark-conditioned trajectory can still diverge from the evidence-supported trajectory. 
In addition, our evaluation is conducted in a controlled RAG setting with selected fact types, and may not cover all forms of factuality errors in open-ended long-form generation. 
Future work should study more diverse domains, longer outputs, human-validated factuality evaluation, and watermark objectives that explicitly incorporate factuality-preserving constraints.

\end{document}